# Global mapping of fragmented rocks on the Moon with a neural network: Implications for the failure mode of rocks on airless surfaces

Ottaviano Rüsch[1], Valentin T. Bickel[2,3]

Running title: Global mapping of fragmented lunar rocks

Affiliations:
1: Institut für Planetologie, Westfäelische Wilhelm Universität Münster, Münster, Germany
2: Center for Space and Habitability, University of Bern, Bern, Switzerland
3: ETH Zurich, Zurich, Switzerland

08.12.2022

# Global mapping of fragmented rocks on the Moon with a neural network: Implications for the failure mode of rocks on airless surfaces

## 0 Abstract

It has been recently recognized that the surface of sub-km asteroids in contact with the space environment is not fine-grained regolith but consists of centimeter to meter-scale rocks. Here we aim to understand how the rocky morphology of minor bodies react to the well known space erosion agents on the Moon. We deploy a neural network and map a total of ~130,000 fragmented boulders scattered across the lunar surface and visually identify a dozen different desintegration morphologies corresponding to different failure modes. We find that several fragmented boulder morphologies are equivalent to morphologies observed on asteroid Bennu, suggesting that these morphologies on the Moon and on asteroids are likely not diagnostic of their formation mechanism. Our findings suggest that the boulder fragmentation process is characterized by an internal weakening period with limited morphological signs of damage at rock scale until a sudden highly efficient impact shattering event occurs. In addition, we identify new morphologies such as breccia boulders with an advection-like erosion style. We publicly release the produced fractured boulder catalog along with this paper.

## 1 Introduction

Fracture, fragmentation and fragments size frequency distribution (SFD) resulting from impact-induced failure of solids are aspects relevant for a wide range of studies (e.g., Kun and Herrmann, 1999). In the case of failure of rocks, these aspects inform on rock strength and are studied for engineering applications and geological research (e.g., Hogan et al., 2012; Zhang et al., 2022). The process of collision and fragmentation is key for the evolution of single and population of planetesimals, asteroids, and solid surfaces (e.g., Capaccioni et al., 1986; Housen and Holsapple, 1990; Giblin et al., 1998; Ryan and Melosh, 1998; Tanga et al., 1999; Durda et al., 2007). The role of impact-fragmentation for solid surfaces is increasing in relevance since it is now known that properties of planetary surfaces can deviate from those of the canonical fine-regolith structure. This has motivated the study of the impact and cratering process into targets composed of coarse to very large particle (e.g., Cintala et al., 1999; Güttler et al., 2012; Kadono et al., 2019; Raducan et al., 2022; Ormö et al., 2022). As particle size increases, the impact kinetic energy used for the particle shattering increases, until there is no energy left for crater excavation: in such case only a shattered or pit-hosting particle (boulder) is left (e.g., Durda et al., 2011) that effectively armor and protect the surface from excavation. For such surfaces, where particles are effectively multi-crystals rocks and boulders, the boulder shattering process, and thus the boulder failure mode, becomes relevant. Here we use the term boulder following the definition of Dutro et al. (1989), i.e., size >25.6 cm, and not that more strict definition of size in the range 0.1-1 m (Bruno and Ruban, 2017) used elsewhere (e.g., Ruesch et al., 2020). There are few opportunities to observe the shattering process in act at the appropriate conditions or to observe the outcomes of naturally occurring fragmentation. Fragmented lunar boulders represent the opportunity to observe the outcomes of such natural events.

Instances of rock failures due to meteoroid bombardment have been observed on the Moon using orbital images acquired by the Lunar Reconnaissance Orbiter Narrow Angle Camera (LROC/NAC) (Ruesch et al., 2020). The exact impact conditions (impactor number, angle, mass, size and velocity) for each instance of boulder shattering is not known, only probability functions are available. Nevertheless, the very slow geological erosion on the Moon, with respect to the one on Earth, allows observations of the pristine morphology (e.g., fragments position and shape) after failure, even if the shattering event leading to failure occurred several tens of million years ago. The study of Ruesch et al. (2020) investigated about ~1000 blocks of size above ~10 m all located around four large impact craters. It demonstrated that the majority of instances of shattered rocks are due to meteoroid bombardment and that the morphology of failed rocks, despite being highly variable, does not vary with surface exposure time.

Several questions follow-up from that study. Are the previously identified morphologies representative of boulder failure modes across the surface of the Moon? Do the failure modes vary selenographically? How do the lunar boulder failure modes compare to recent observations of failure modes on other bodies, specifically on near Earth asteroids Itokawa (e.g., Nakamura et al., 2008), Bennu (e.g., Molaro et al., 2020; Delbo et al., 2022), and Ryugu (e.g., Sugimoto et al., 2021)?

This study uses a neural network on LROC/NAC images (Robinson et al., 2010) in order to search for fragmented blocks on a quasi-global scale. The neural network was loosely trained on images that cover the full range of known block degradation states, including boulders with single fracture, heavily fractured boulders, as well as catastrophically shattered boulders, specifically excluding large, intact boulders, however. The resulting network is capable of identifying fractured media from the meter- to hecto-meter scale.

## 2 Method

Our methodological approach consists of three main steps: 1) neural network-driven detection & mapping, 2) candidate clustering, and 3) visual inspection & review.

**Neural network**

We utilized a convolutional neural network architecture called RetinaNet (Lin et al., 2017) with a ResNet50 backbone that has been successfully deployed for a series of lunar, machine learning mapping-related studies in the past (Bickel et al., 2020a; Bickel et al., 2021a; Bickel et al., 2018). A total of 151 positive and 67 negative training labels were manually created by a human operator following previously established procedures (Bickel et al., 2020b; Bickel et al., 2021b). Training labels intentionally included a wide range of boulder disintegration states (Fig. 1), ranging from boulders with single fractures to completely disintegrated boulders; intact boulders were specifically excluded, i.e., included as negative training labels. The neural network was trained over a total of 75 epochs, utilizing label augmentation, such as image flipping, rotation, re-sampling, shearing and radiometric adjustments (brightness and contrast) to increase network robustness.

We note that the nature, distribution, physical appearance, and frequency of fractured boulders and outcrops on the Moon is not known exactly, which means that any supervised machine learning approach (like ours) is inherently biased. In other words, our training dataset can - by definition - not be representative of the actual distribution of fractured boulders and outcrops on the Moon. This is why we did not perform any testing of the trained network - the motivation of this work was to identify and study a very broad and unknown range of fractured features on the lunar surface, not a very specific, known feature. We purposely designed and trained the neural network to have a very wide perception, enabling the detection and categorization of an unknown number and types of fractured features. This implies that the results of this study are of a qualitative nature - our incomplete understanding of the performance of the network means that no meaningful quantitative conclusions must be drawn.

We deployed the trained neural network in an existing LROC/NAC processing pipeline (Bickel et al., 2022; Bickel et al., 2020a). The output of this pipeline includes a list of fractured feature candidate detections (here called "candidates"), with relevant information regarding location and size of features, as well as candidate preview patches, i.e., full-resolution thumbnails (~100 by ~100 pixels on average) of each candidate detection that are useful for review and science analysis. The deployment of the neural network was limited to images with incidence angles between 10 and 60° to avoid steep slopes to be shadowed (observational bias). This effectively prevented the neural network from scanning regions beyond 60° N and S.

**Clustering**

Each candidate identified objects of very different geological nature, predominantly geomorphic features that are represented by strong reflectance contrast(s) with a certain spatial pattern, specifically linear and quasi-linear patterns (as the network was trained on). In addition, the network identified visually similar objects that are not geologically related to fractured boulders, such as rocky craters and fracture sets in impact melt deposits. The sheer number and wide range of represented geomorphic features makes a systematic analysis difficult. To facilitate human review and analysis, we used an unsupervised (saliency-based) clustering approach to sort all candidates in clusters that share certain visual characteristics, adopting the methodology developed and used by Bickel et al. (2022). Specifically, we use a pre-trained (ImageNet) VGG16 to identify relevant features in the candidate images and a k-means approach to subsequently cluster them into 15 classes.

**Visual inspection**

Some of the clusters consisted predominantly of features not related to fractured boulders; we removed those clusters from further analysis. In each of the feature-positive, selected clusters we visually searched for recurring patterns in the morphology of the cluster of fragments. Practically, the analysis consisted in a rapid photo-geological interpretation of the thumbnails, i.e., i) identification of discrete surfaces within the thumbnail, ii) association of these surfaces to geological units (rocks, soils) and iii) mental reconstruction of the three-dimensional rocks and soils units (e.g., Wilhelms et al., 1987). Attention was paid to the number, size and shape of the fragments, the position of the fragments with respect to each other, and the presence of cracks. To characterize the fragments we use the ratio of the area of the largest fragment to the area of the parent (Alf/Ap), with the latter approximated by the area containing the group of rocks. Challenges of this dataset, and of optical images in general, is the lack of topographic information and the presence of shadows that obliterate areal fractions of the scene. In order to avoid misinterpretation of morphologies only morphologic types occurring at least twice are reported. Use of multiple images acquired at different local times could help to further improve the interpretation of the morphologies.

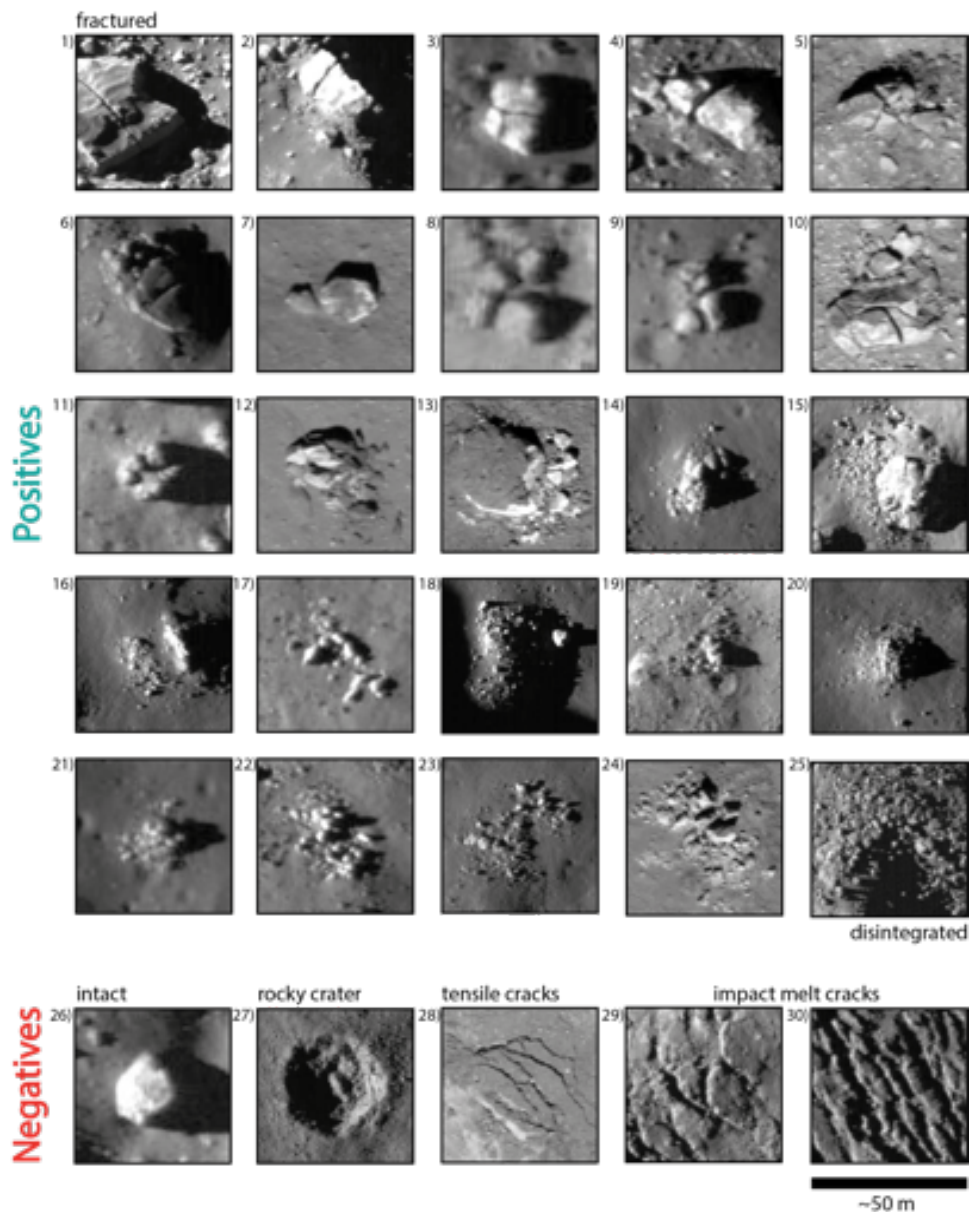

Fig. 1. Examples of positive and negative labels used for the training of the neural network. Note how the full range of boulder disintegration is covered, from fractured all the way to catastrophically shattered (1 through 25). Negative labels include, among other things, intact boulders, craters, and linear features unrelated to boulder disintegration (26 through 30). Image credits to NASA/LROC/GSFC/ASU.

## 3 Results

## 3.1 Overview

The machine learning algorithm identified a total of 187,787 fractured feature candidates, which were clustered into 15 individual clusters. After visual inspection, we found that 14 clusters predominantly contain features that are directly related to boulder fragmentation; 1 cluster predominantly contains features that are unrelated.

Of the 15 clusters, 14 are considered relevant for further investigation with a total of 159,442 features. Cluster 8 was disregarded. We note that clusters 0, 1, 3, 10, and 14 contain a substantial portion of crater-related features (containing 72,929 features), thus, are described as "heterogeneous". Due to the overwhelming number of candidates in each cluster, only about 50% of the detections in each relevant cluster are visually analyzed. A dozen morphological types are identified and are described in the section 3.3. We note that several of these types are morphological end-members and that a continuum of morphologies is observed in between these endmembers. The discarded cluster 8 predominantly contains fresh rocky and/or concentric craters. We include a summary table with cluster ID and overall representative features below (Table n.). The produced catalog and candidate preview patches are available online, free of charge, here (interim link, pw is 'fracture'): https://polybox.ethz.ch/index.php/s/PV1JNydBKLtQfYO

Table 1. Summary of all produced clusters, including archetype description, number of features, mean estimated diameter (using the approach described by Bickel et al., 2018), and an exemplary NAC crop-out; note that the given example for heterogeneous clusters is not representative of the full range of features. Image credits to NASA/LROC/GSFC/ASU.

| Cluster ID | Dominant archetype | No. of features | Mean est. diameter | Example |
|---|---|---|---|---|
| 0 | Heterogeneous cluster, large- to small-sized fractured boulders mixed with many fresh and rocky craters | 14,206 | 8.9 m | 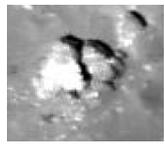 |
| 1 | Heterogeneous cluster, mid- to small-sized fractured boulders with many fresh, rocky, and concentric craters | 10,614 | 15.5 m | 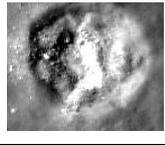 |
| 2 | Homogeneous cluster, large- to mid-sized fractured boulders | 6,852 | 10.9 m | 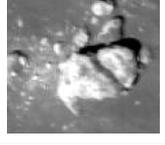 |
| 3 | Heterogeneous cluster, large- to small-sized fractured boulders mixed with many fresh and rocky craters | 17,863 | 11.1 m | 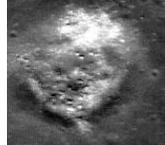 |
| 4 | Homogeneous cluster, mid- to small-sized fractured boulders | 14,103 | 8.2 m | 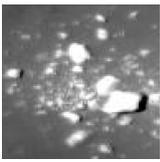 |
| 5 | Homogeneous cluster, large- to mid-sized fractured boulders | 13,084 | 7.8 m | 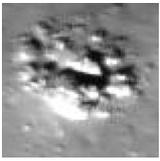 |
| 6 | Homogeneous cluster, large- to mid-sized fractured boulders | 11,025 | 8.2 m | 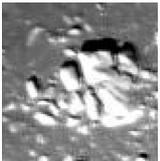 |
| 7 | Homogeneous cluster, mid- to small-sized fractured boulders mixed with some fresh craters | 11,288 | 13.0 m | 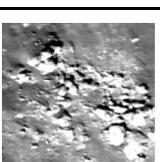 |
| 8 | Homogeneous cluster, fresh, concentric, and rocky craters | 28,345 | 7.9 m | 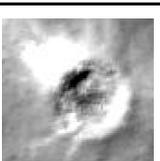 |
| 9 | Homogeneous cluster, mid- to small-sized fractured boulders | 8,009 | 9.9 m | 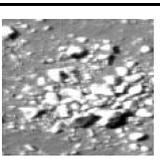 |
| 10 | Heterogeneous cluster, large- to small-sized fractured boulders mixed with many fresh rocky craters and some intact boulders | 18,650 | 8.1 m | 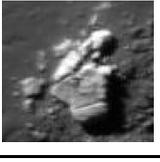 |
| 11 | Homogeneous cluster, mid- to small-sized fractured boulders | 8,765 | 8.2 m | 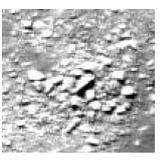 |
| 12 | Homogeneous cluster, large- to small-sized fractured boulders | 6,212 | 10.3 m | 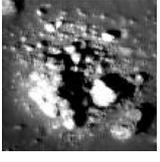 |
| 13 | Homogeneous cluster, mid- to small-sized fractured boulders | 7,175 | 8.3 m | 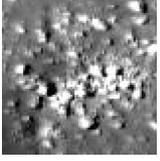 |

| 14 | Heterogeneous cluster, large- to small-sized fractured boulders mixed with some fresh, rocky, and concentric craters | 11,596 | 10.2 m | 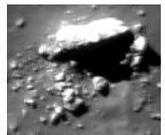 |

## 3.2 Maps

We produced a series of maps to visualize the spatial distribution and frequency of a number of particularly interesting clusters. Fig. 2 shows clusters 2 (large fractured boulders), 8 (predominantly impact craters), and 13 (small fractured boulders); we note that clusters which predominantly contain impact craters appear to be scattered randomly, while clusters which predominantly contain fractured boulders appear to be systematically located. Globally distinct hotspots (spatial clusters) of fractured boulders are predominantly located in and around craters. Interestingly, fractured boulder hotspots are located in craters of all ages, ranging from Copernican- to Imbrian-aged craters. Hotspots of fractured boulders of different clusters - i.e., with different morphological properties - are not necessarily co-located; for example, Tycho features a prominent cluster of large fractured boulders, but not of small clustered boulders. Similarly, Mare Serenitatis features a significantly increased number of small fractured boulders, but an apparent lack of large fractured boulders (Fig. 2). It is important to keep in mind that the used neural network suffers from a fundamental bias which might result in potentially misleading maps.

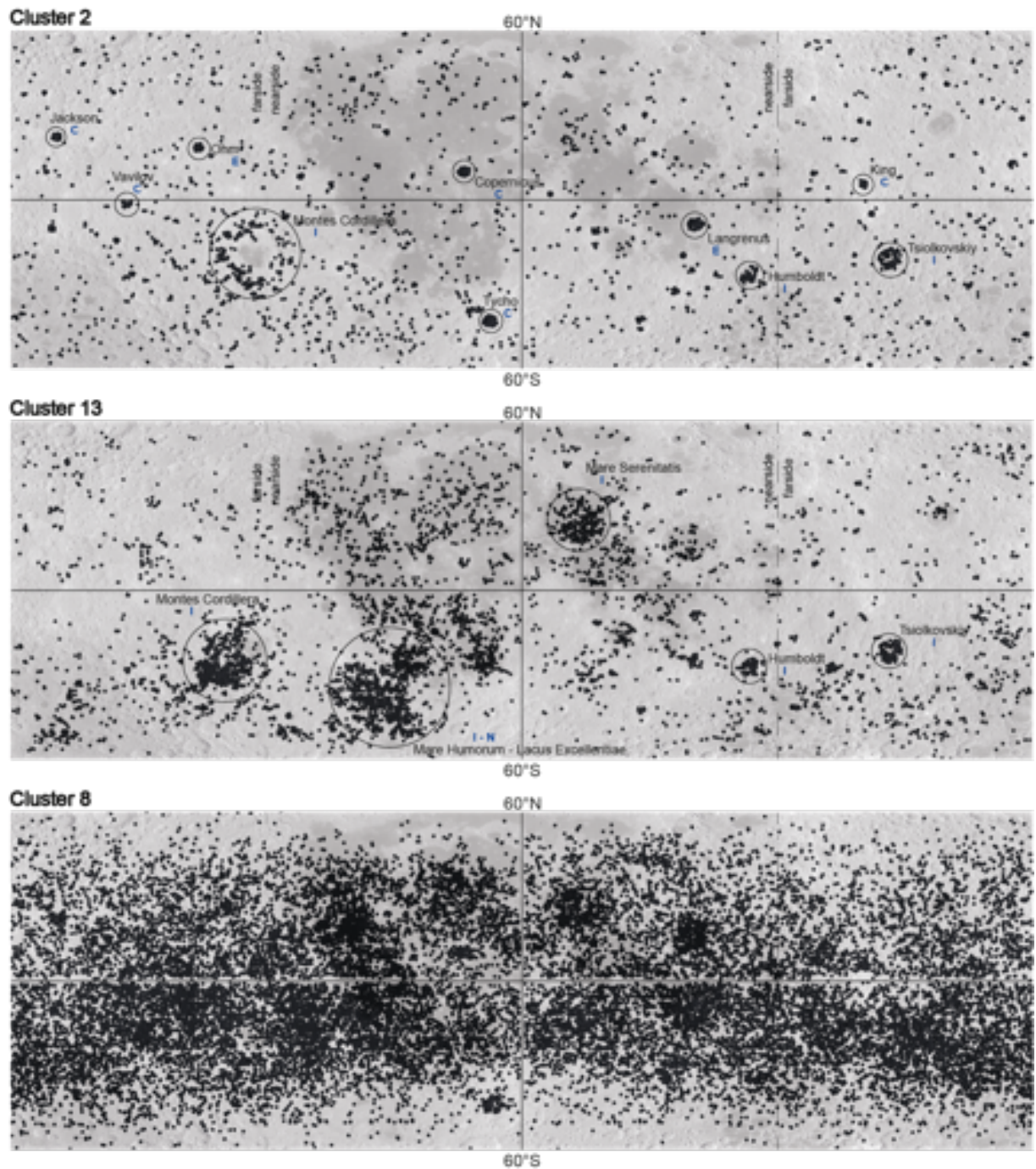

Fig. 2. Spatial distribution of clusters 2 (predominantly large fractured boulders), 13 (predominantly small fractured boulders), and 8 (predominantly impact craters), and 13 (predominantly small fractured boulders), black dots. A few selected, globally distinct hotspots are highlighted by black circles; host location age indicated by a single letter: C = Copernican, E = Eratosthenian, I = Imbrian, N = Nectarian. WAC global mosaic in the background; WAC image credits to NASA/LROC/GSFC/ASU.

### 3.3 Morphological types

This section contains detailed qualitative descriptions of the fractured boulder morphologies we observed throughout the dataset. These morphological types are found in one or several of the identified clusters.

*Cavity (Fig 3a,b):* These instances are characterized by an intact boulder, i.e., without fractures, with one of the facets presenting a fracture plane curvilinear in planar view. The fracture plane is identifiable because it is partly in shadow. *Interpretation:* The fracture plane represents a cavity into the boulder. In one case fragments the size of the cavity are located in proximity of the cavity near the boulder and might represent excavated material from the cavity. A similar morphology has been reported on asteroid Bennu (Molaro et al., 2020).

*Interrupted near-horizontal spalling (Fig 3c,d)*: These boulders present narrow linear shadows that partly, not completely, cross a facet. There is no spatial displacement of the illuminated rock material on both sides of the linear shadow. *Interpretation*: The topography responsible for such shadow configuration can consist of a step-like relief where the topographically lower side is in shadow. Step-like relief can form by interrupted spallation when the energy necessary for spalling an entire slab is insufficient and the crack changes direction toward the nearest surface. The fragmented removed by the spall is not visible because i) it was displaced outside the frame, ii) it has been shattered in small unrecognizable pieces, or iii) occurred before the emplacement of the boulder at the current location. Similar morphology and interpretation has been reported by Nakamura et al., (2008). Unlike for the other morphologies, the formation of the features for this type can have occurred during formation of the block itself rather than post-emplacement. The presence of a vertical crack (deep incision) instead of an interrupted spall is excluded because the sun-facing wall of the crack observed in other morphologies (see interrupted fracturing) is not identifiable.

*Interrupted or deviated fracturing (Fig 3e,f):* These boulders present narrow linear or zig-zag shadows that partly or completely cross a facet. The illuminated sides juxtaposing the shadow do not have the same brightness and one side can be as bright as the brightest area of the image. *Interpretation:* The juxtaposing of shadow with relatively high brightness indicates a crack topography with a trough sufficiently deep and with steep wall. The illuminated steep wall can be responsible for the relatively high brightness next to the shadow. Crack depth and width is likely increased by a displacement of the newly created fragments. Zig-Zag pattern can represent the deviation of the original crack propagation direction toward the nearest surface. Here the crack appears to propagate through the boulder, whereas in the interrupted spalling type, the crack affected the boulder only superficially.

*Shear failure* (Fig. 3g,h)*:* Boulder presenting a curved narrow shadow of variable width, juxtaposed to a relatively bright illuminated area of rock. The narrow shadow separates the parent boulder in two fragments. The smaller fragment is not as small as in the case of near-vertical spalling. *Interpretation:* The curved pattern of the narrow shadow indicates that the dipping of the crack is not vertical but tilted, implying shear failure along a single plane that led to the partial superposition of two large fragments. Shear failure is a typical failure mode observed, for example, in compression tests (e.g., Basu et al., 2013).

*Near-vertical spalling (Fig. 3i-l):* In planar view a narrow shadow runs near and parallel to the edge of a facet. In some places along its length, the shadow can increase in width and be juxtaposed to an illuminated boulder area of high brightness. *Interpretation:* These are examples of spalling of rocks where the largest surface of the spalled flake is oriented vertically. The flake can either be mostly attached to the parent block or located slightly away from it. This morphology bears strong similarity with exfoliation morphology reported on Bennu (Molaro et al., 2020). The thickness of the proposed spalled flakes on Bennu relative to their parent boulder is much smaller than what is observed here. We note, however, that the resolution of the images for the Moon does not allow to detect thinner flakes.

*Axial failure (Fig. 3m-p):* Boulder with one or more narrow linear shadows of constant width running through an entire facet. In planar view the contact between multiple shadows occurs predominantly at either an approximately right angle (n=5) or at an angle in the range 30-40 degrees (n=6). *Interpretation:* Shadows resulting from near-vertical fractures. In some cases fractures cut through the entire boulder height and the resulting fragments are slightly displaced. As noted previously in Matsui et al. (1982), these fractures resemble axial failures observed in static uniaxial compression tests (e.g., Basu et al., 2013) where a rock is splitted along one of its axes. This morphological type on the Moon was already reported in Ruesch et al. (2020) where it was suggested to be due to thermal stresses due to the non-random orientation of cracks. As discussed further below, the new instances detected in this study suggest that single and multiple axial failures could also be part of a continuum of morphologies resulting from a wide range of imparted impact energy.

*Concentric fracturing (Fig. 4a-c):* This type of morphology, as well as the followings, is characterized by more than two fragments. The boundary of the central fragment presents a semi-circular shadow. The configuration of the shadow bears resemblance with the shear failure type. In this type, however, more than two fragments are present and the central fragment can be rounded in planar view. *Interpretation:* The only resemblance with literature of solid fragmentation is the presence of a fragment at a central location, i.e., surrounded by fragments on all side, like in the so-called core shattering events (Fujiwara et al., 1989). Core shattering fragmentation can develop shell-like fractures surrounding the core (e.g., Fujiwara and Tsukamoto, 1980; Durda et al., 2015) similar to the concentric fracturing observed here. In our examples the characteristic is that the central fragment is round and that is not necessarily the largest fragment.

*Core (Fig. 4d-f):* This morphology is composed of a group of several fragments. One fragment has all its sides surrounded by other fragments. This fragment is located at a central location, i.e., other fragments are distributed around it in all directions and can, in some cases, be the largest. The number of relatively small fragments is variable. *Interpretation:* The central fragment is a core, i.e., a fragment whose facets are not the original outer surface of the parent block. This type of fragmentation can be associated with confidence to the core shattering type described by, e.g., Fujiwara et al. (1989), Fujiwara and Tsukamoto, (1980), and Durda et al. (2015). The number of very small fragments is likely due to the age of the shattering event, i.e., high for recent events and low for old events (Rüsch et al., 2022). We cannot exclude, however, that shattering events producing only very few small fragments occur. Several instances of this group are very similar to the "concentric fracturing" type, implying a continuum of morphologies.

*No core (Fig. 4g-i):* This morphology is composed of several fragments. The positions of the largest fragments form a circular feature, i.e., it encircles an area lacking relatively large fragments. The center of the feature presents relatively very small or no fragments at all, effectively a "stone circle" landform. The number of relatively small fragments surrounding the large fragments is highly variable. *Interpretation:* This type is similar to the "core" type without the central fragment (the core). See the subsection "cluster" for a discussion of the very small fragments. An instance similar to this type is the Loong Rock (Outer Fence) group of fragments at the Chang'E-3 landing site (Li et al., 2015; Li and Wu, 2018; Di et al., 2016).

*Cones (Fig. 4j-l):* This morphology is a group of fragments with one or more fragments displaying a triangular and or elongated shape in planar view (cone-like) and with the apex located toward the center of the cluster. Unfortunately not the entire three-dimensional shape of these elongated fragments can be discerned, thus while one aspect ratio is known to be very high (a>>c) we have no information on the other (relationship a,b). *Interpretation:* This morphology corresponds to the well known "cone-type" fragmentation reported in Fujiwara et al. (1989), Matsui et al. (1982), and in Giblin et al. (1994; 1998). Elongated or flattened fragments are often formed on the outer target surface, especially in the cratering regime (Fujiwara et al., 1989; Nakamura et al., 2008; Walker et al., 2013). Here, however, the presence of these fragments near the center of the cluster and not on its edge, as well as their rather straight and not curvilinear form, is consistent with the formation of such fragments inside the parent target, as found by Durda et al. (2015). This type of morphology is consistent with dominant radial fracturing.

*Cluster (Fig. 4m-o):* Group of fragments with the area of the largest fragment much smaller than the area covered by the group ($A_{lf}/A_p<0.1$). The number of fragments is >50. The cluster can have a concentration of fragments at its center or along a circle. Fragments can form rays radiating from the center of the cluster. The surface in between fragments can have a higher reflectance than the background surface away from the cluster. *Interpretation:* The ratio $A_{lf}/A_p$ can be used as a proxy for the ratio of the mass of the largest fragment to the mass of the parent target (e.g., Ruesch et al., 2020). Based on this parameter the morphology corresponds to highly catastrophic events where the mass of the largest fragment is much smaller (<0.1) than the mass of the parent.

*Bright zone (Fig. 5a-c):* Presence of a single high reflectance patch. Where the patch is relatively large, it is centered on a boulder fracture or a fragment edge. The reflectance of the patch is higher toward its center. The boundary of the patch is gradational and uncorrelated to fragment edges or other topography. There is no evident association to a specific morphological type of fractured or fragmented boulder. *Interpretation:* High reflectance due to unweathered rock and soil surfaces typical for very recent exposures (e.g., Pieters and Noble, 2016). The exposure of fresh material is likely due to a meteoroid impact (e.g., Hörz et al., 1975; Ruesch, 2020) with the patch corresponding to the spall zone on a rock and to adjacent ejecta material.

Since the patches are observed to spatially coincide with a fracture or with the edges of juxtaposed fragments, they are direct evidence that a meteoroid event is responsible for the fracturing and fragmentation. This same co-occurrence of a fragmented boulder and brightening can be noted at White Rocks of station C1, Apollo 14 (Ruesch, 2020), possibly enhanced by the high albedo of the anorthitic petrology. Very similar white spots on blocks on Itokawa have been interpreted as resulting from meteoroid impacts (Nakamura et al., 2008).

*Plate-like fragments (Fig. 5d,e):* Relatively rare assemblage of fragments with one of the largest fragments of very high reflectance and with an abrupt transition between illuminated high reflectance area and shadow. The bright fragment can display a relatively high ratio of shadow length to rock width. *Interpretation:* This configuration suggests the presence of a plate-like fragment, with an inclined (~40-60 deg) side. The spatial configuration of illuminated and shadowed surfaces indicate that the fragment forms an overhang. This overhang configuration is likely the result of the plate-like shape and the rather vertical position of the fragment. The high reflectance surface is due to favorable scattering from the steeply inclined fragment side. Flattened fragments are rather common in solid fragmentation as noted in literature (e.g., Durda et al., 2015) and in the morphological types described so far.

*Partially buried (putative) (Fig. 5f,g):* These are boulders or fragments partially or completely inside a shallow depression. The shape of the depression can be circular or irregular. *Interpretation:* the co-occurrence of fragments in or near a crater-like depression can be a coincidence. Alternatively, it could represent impact-induced fragmentation of a partially buried boulder. As demonstrated experimentally, impact on a partially buried rock leads only to minimal damage and is associated with the formation of a main crater of circular to irregular shape and several smaller craters (Durda et al., 2011).

*Breccia (Fig. 6):* This is a morphology that does not present spatially resolved fractures or detached fragments and nevertheless show signs of disaggregation. These instances are characterized by heterogeneous reflectance of boulder facets together with relatively small shadow on top of the main boulder facet. The heterogeneous reflectance forms several patches with sharp boundaries. Some of the patches have a shadow at their boundary. The boundary of the boulder is highly irregular and leads to partial enclosure of the juxtaposed soil (Fig. 6a). The partial enclosures can extend into linear and curvilinear features on the facet of the boulder (Fig. 6b,f). These boulders are found on the rim of old (>~100 Myr) craters as part of a mature population of ejecta blocks. *Interpretation:* The patches represent clasts within a matrix and the boulder thus constitute a breccia. The partial enclosures and linear to curvilinear features are evidence for incision and trough formation into the boulder where the rheology, probably the matrix, is more susceptible to the micrometeoroid abrasion. The clasts are more resistant to micrometeoroid and develop positive reliefs. The topography can be described in terms of advection, i.e, removal of material with retreat and preservation of slope, and contrasts with the diffusion-like topography typical for monolithic boulder (e.g., Rüsch and Wöhler, 2022) where removal of material is associated to a change in slope. Probably this morphology develops slowly and is only evident in old (>100 Myr) survivor boulders. The hummocky texture and presence of bright clasts is very similar to the boulder containing intra-boulder bright clasts on Ryugu (Sugimoto et al., 2021) and Bennu (DellaGiustina et al., 2021).

*Boulder field with pattern (Fig. 7):* Field of non-fractured boulders displaying a radial pattern, e.g., broad and short rays, and a central area lacking boulders. The spatial density of boulders is very high, i.e., boulders can be in contact with each other, and the boulders themself show no sign of fragmentation. *Interpretation:* Morphology resulting from an impact into a fine-grained target (i.e., regolith) containing large inclusions (i.e., pre-existing boulders or fractured bedrock) and different from that of an impact into a target composed of only fine-grained or large particles (e.g., Güttler et al., 2012). The spatially inhomogeneous distribution of boulders around the impact site, i.e., the rays and the pattern, are consistent with laboratory and numerical simulations of impact into a target with a wide range of particle sizes without a dominant particle size (e.g., Kadono et al., 2019; Kadono et al., 2020; Ormö et al., 2022). The observed radial pattern is itself the result of filaments structure in the ejecta curtain (e.g., Kadono et al., 2022). On the Moon, this morphology indicates the presence of a subsurface rich in boulders or of an underlying pre-fractured bedrock.

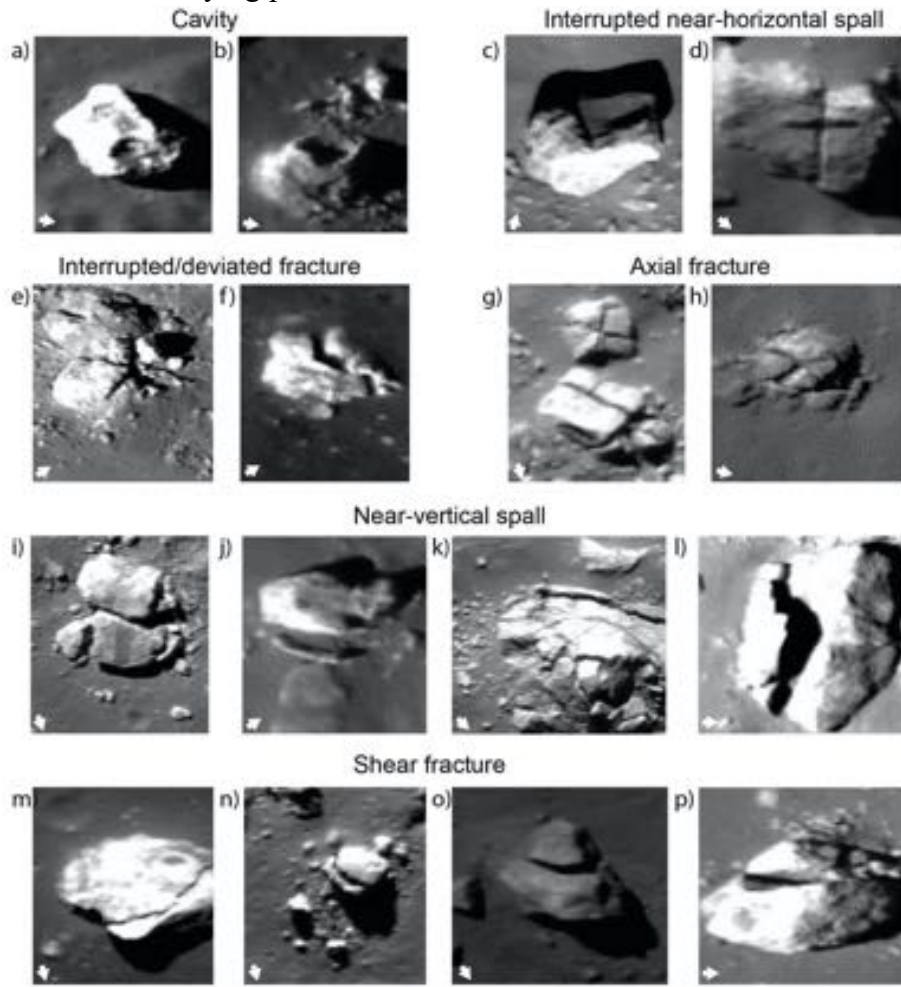

Fig. 3. Examples of detections for different morphological types. Large arrow denotes illumination direction. Panels width varies in the range 30-60 m. a) M174889254LC, (b) M176602625RC, (c), M124999473RC, (d) M1251998743RC, (e) M152702349RC, (f) M126595195LC, (g) M1162548409LC, (h) M1221615756RC, (i)

M1222952209LC, (j) M112671946LC, (k) M108292443LC, (l) M1296942537RC, (m) M170905869LC, (n) M172460917LC, (o) M157364573RC, (p) M175408129RC. Image credits to NASA/LROC/GSFC/ASU.

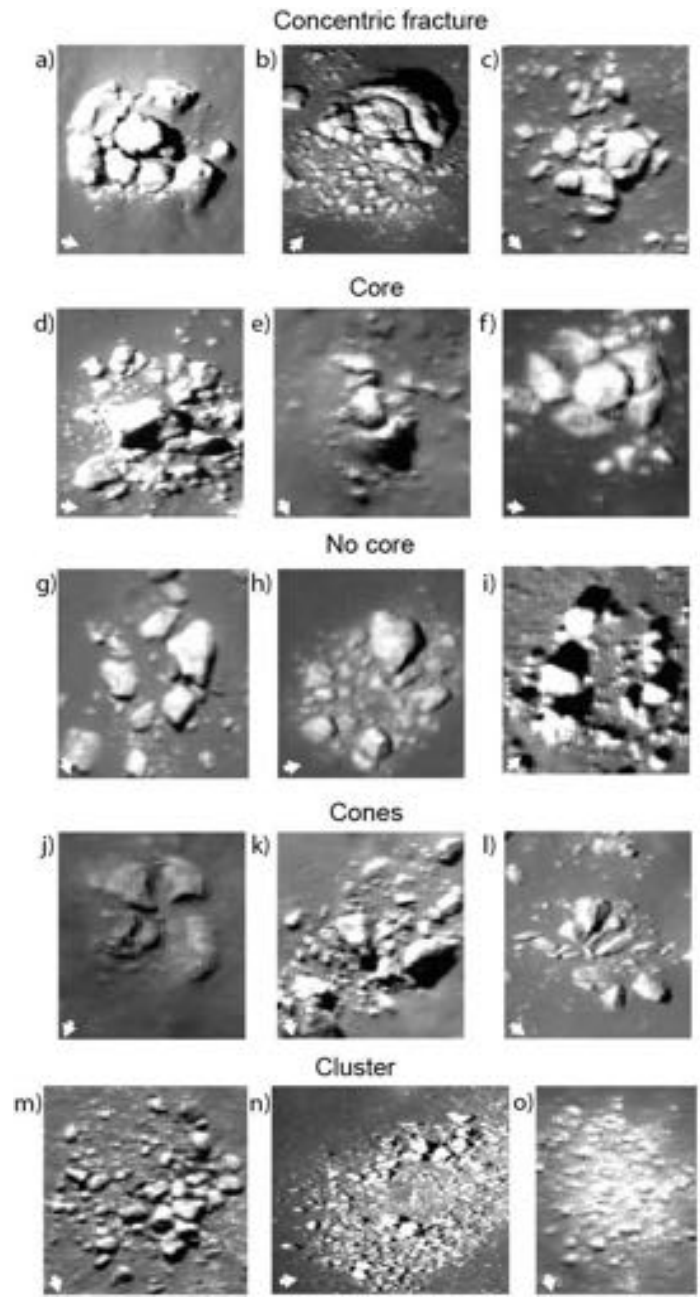

Fig. 4. Examples of detections for different morphological types. Large arrow denotes illumination direction. Panels width varies in the range 30-60 m. (a) M1356624061RC, (b) M169845868RC, (c) M1193717676LC, (d) M127240244LC, (e) M110656476LC, (f) M108203484RC ,(g) M1193830064LC, (h) M110560726LC, (i) M1327990586LC, (j) M108366721LC, (k) M139029855LC, (l) M169283456RC, (m) M109780793RC, (n) M168523403LC, (o) M1314055387RC. Image credits to NASA/LROC/GSFC/ASU.

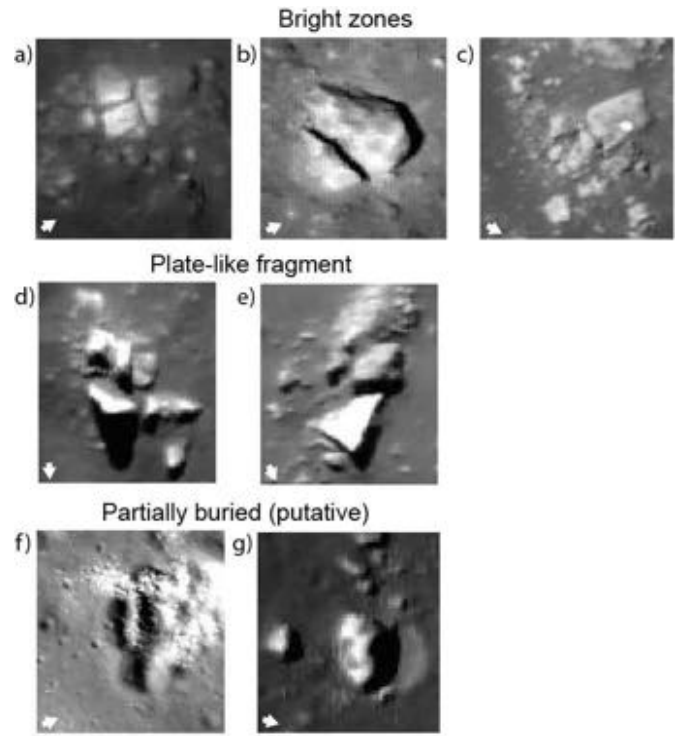

Fig. 5. Examples of detections for different morphological types. Large arrow denotes illumination direction. Panels width varies in the range 30-60 m. (a) M110560726RC, (b) M109895309LC, (c) M110526893RC, (d) M1299030725RC, (e) M110649693RC, (f) M1280467964LC, (g) M1130300625RC. Image credits to NASA/LROC/GSFC/ASU.

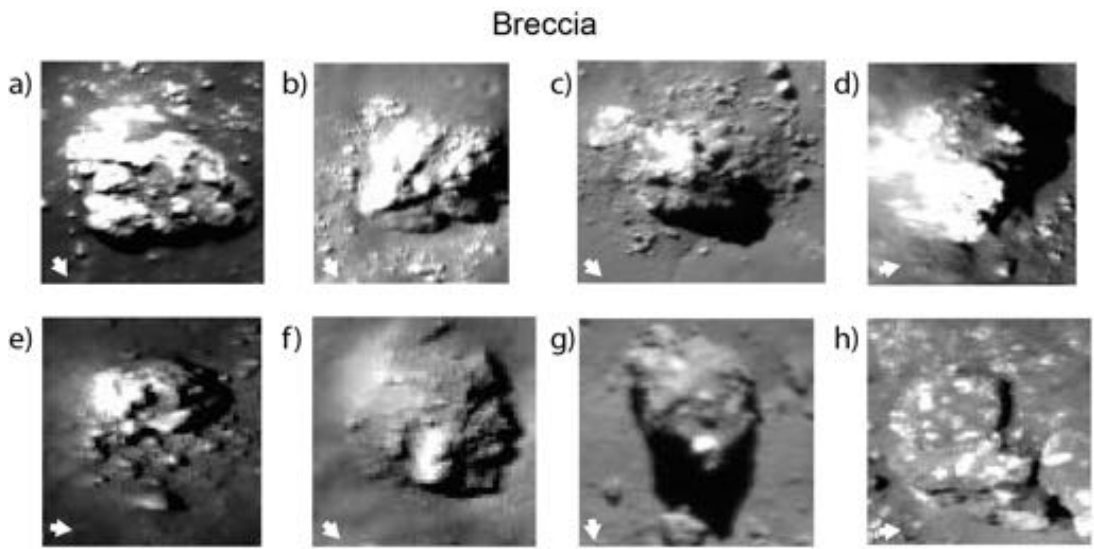

Fig. 6. Examples of detections for the "breccia" morphological type. Large arrow denotes illumination direction. Panels width varies in the range 30-60 m. (a) M110126713LC, (b) M1325718594RC, (c) M111668133RC, (d) M137278395RC, (e) M168869503RC, (f) M107809965LC, (g) M1343664677RC, (h) M127070409LC. Image credits to NASA/LROC/GSFC/ASU.

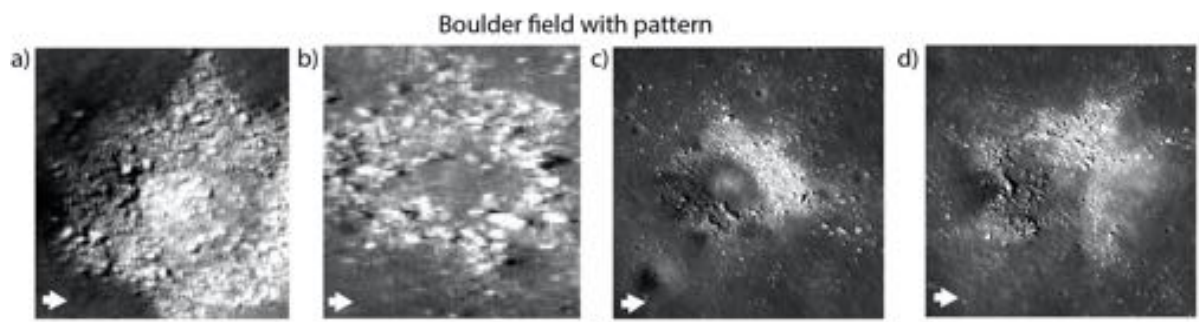

Fig. 7. Examples of detections for "boulders field with pattern". Large arrow denotes illumination direction. Width of panels (a, b) is 50 m. The view for the detections in (c,d) has been enlarged to 300 m width. (a) M1221580467RC, (b) M168442030LC, (c,d) M137420922LC. Image credits to NASA/LROC/GSFC/ASU.

**Discussion**

The foundation of this work is a catalog of more than 180,000 fractured feature candidates detected and mapped by a neural network. As the true variety of boulder fragmentation morphologies on the Moon is inherently unknown, any supervised machine learning-driven mapping approach includes one or many unknown and potentially significant bias(es). We employed a best effort approach to make the neural network as robust as possible to the full range of known morphologies, which turned out to be successful - to an unknown extent - as a number of previously unknown fragmentation morphologies, that were not included in the training data (see e.g., Fig. 3a,b, 6, 7), were identified. It is important to note that it remains unclear whether the catalog and descriptions presented in this work are complete. Any quantitative conclusion drawn from this work might suffer from potentially significant bias(es).

In general, morphologies in Cluster 2 as defined by the unsupervised clustering (Section 3.3, Table 1) can be interpreted in terms long-lasting landforms such as large fractured survivor blocks belonging to mature (>50-100 Myr) ejecta block fields or to fractured exposed bedrock, e.g., on central peaks of large craters. Morphologies in Cluster 13 represent short-lived landforms such as smaller fragmented blocks in immature ejecta block fields. For such younger immature ejecta block fields the larger blocks are not yet fractured or, due to the small size of the crater, are not formed. A comparison between the maps of cluster 2 and 13 with the rock abundance map retrieved from thermal observations by Diviner (Bandfield et al., 2011) demonstrates that rock-rich regions can be very different. For example, regions of similar Diviner rock content such as Copernicus and Mare Humorum correspond to large (~20 m) fractured blocks and to small fragments (<10 m) of shattered blocks, respectively. In other words, different size-frequency distribution of rocks and different spatial configuration can produce the same rock abundance measured by Diviner. These differences could be used to improve the estimate of rock abundance and surface properties from the Diviner data in future analyses.

Many of the morphologies detected in the visual survey (Section 3.2) have been described in previous work on rock failure (e.g., Basu et al., 2013; Zhang et al., 2022) or shattering (e.g., Fujiwara et al., 1989; Nakamura et al., 2008; Durda et al., 2015), although it is the first time that these morphologies are observed for very large target sizes (10-50 m) and for the velocity regime of ~10-20 km/s. The similarity of morphological types for impact-induced fragmentation occurring for a wide range of target size (cm to tens of m) is consistent with scale invariant fractal flaw distribution (e.g., Turcotte, 1997; Housen and Holsapple, 1999). Several of the identified morphologies can be grouped in two overarching classes characterized by the flaw geometry: concentric and radial flaws. Based on the geometry of the flaws we propose that "concentric fracturing" and "near-vertical spalling" are lower energy modes of the "core" type, characterized by concentric flaws. Instances of radial fracturing present in the "axial failure" type are lower energy mode of the "cones" and "no core" types mode, characterized by radial flaws. Importantly, we find that these two flaw geometries, exemplified by the types "cones" and "core" already described in Fujiwara et al. (1989), do not appear to depend on the imparted energy inferred as the ratio size largest fragment/size parent. For example, there are cones mode with the largest fragments having a size of about 1/3 the parent. There are also core modes with the largest fragment (only one) with a size of about 1/3 the parent. This means that for about the same size of the largest fragment, either the cones or core mode could develop. This duality can be observed for high energy events as well. For example, in two instances of the "cluster" type (Fig. 4m,n,o and reproduced in Fig. 8) the ratio Alf/Ap is similar. However, in Fig. 3m,o the largest fragments are at the center, whereas in Fig. 3n, the largest fragments distributed in a near circular pattern. The difference might be due to the impactor property, i.e., difference in velocity or mass for the same imparted energy, and/or target property. This is consistent with suggestions in Giblin et al., (1998) where cone type was generally associated to large and low velocity impactors and core type to small and high energy impactors. It is likely that while our observations suggest that one of the two flaw geometries can dominate fragmentation, both types are involved in the process.

There are additional associations that can be established between several of the morphological types. Figure 8 illustrates the fragmentation energy and temporal relationships between morphologies, with solid and dashed lines, respectively. The relationship based on the fragmentation energy described in the preceding paragraph is exemplified by the close-up images on line "a" and "b". The fragmentation energy is qualitatively inferred from the ratio Alf/Ap. The temporal relationship between any type of morphology is exemplified by the dashed lines. If small fragments are formed at the time of fragmentation, usually in a high energy event, their number decreases with time such that only large fragments will remain after a few tens of million years (Rüsch et al., 2022). In addition, very recent events are likely to display a bright surface due to the presence of unresolved fragments (e.g., Marshal et al., 2023) or unweathered particles (e.g., Pieters and Noble, 2016). As the close-up images in Figure 8 illustrates, when the exposure age is more than sufficient to erase all small fragments, there is development of debris apron (fillet) identifiable by a bright halo adjacent to rocks. In the extreme case of a very old exposure age, there is self burial as described in Rüsch and Wöhler (2022). This is only observed for "survivor" boulders that did not undergo fragmentation.

We note that there exists degeneracy on some of these relationships because it is challenging to determine whether a fragmentation event occurred without formation of small fragments, or whether the small fragments have been erased after their formation. In Figure 8 we interpret the absence of small fragments to be due to their erasure. This is however not necessarily true for all instances. In particular mild sub-catastrophic events, with the Alf/Ap>0.5, might have been formed without production of small fragments as shown by examples of fragmented blocks with bright zones.

Finally, we detected several instances testifying for very high fragmentation energy where the center of the cluster-type morphology presents a shallow depression. These detections shown in Figure 8 represent the end of the so-called armoring regime and the start of the cratering regime of the underlying surface. This situation is equivalent to the case for rubble-pile surfaces of asteroids where the impact energy is mostly dissipated by the fragmentation of the first-contact block and little energy is available for crater excavation (e.g., Tatsumi and Sugita, 2018; Bierhaus et al., 2022).

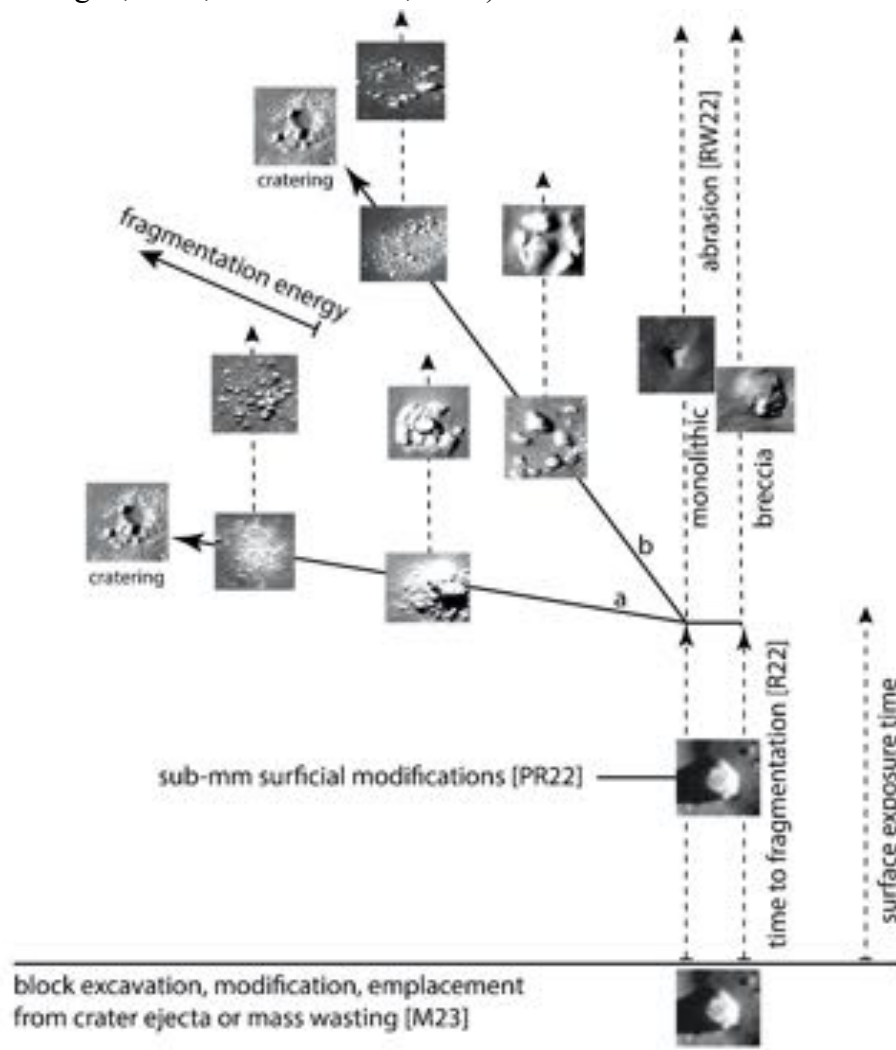

Fig. 8. Summary and relationships of lunar boulder morphologies under erosion. Relationships between morphologies can be described in terms of exposure time after emplacement and impact fragmentation energy. Before fragmentation, changes of morphology are limited to sub-mm surficial modifications. After fragmentation (where it occurs), changes of morphology due to increase in exposure age (connected by dashed lines) correspond to a loss of high reflectance, and/or decrease of small fragment number and/or development of debris apron. Line a, b represents the end-member morphologies of the "core" and "cones" failure type, respectively. Survivor boulders that are not fragmented by a large meteoroid develop debris aprons by abrasion. Specific studies are

M23: Marshal et al., 2023; R22: Rüsch et al., 2022; PR22: Patzek and Rüsch, 2022; RW22: Rüsch and Wöhler, 2022; Image credits to NASA/LROC/GSFC/ASU.

As stated earlier, some boulder morphologies on asteroid Bennu have been ascribed to erosion by thermal stresses (e.g., Molaro et al., 2020; Walsh et al., 2019). We have detected these same morphologies. The morphology on Bennu described as disaggregation in place with disaggregated fragments on and around the parent block (Molaro et al., 2020) correspond to the "localized damage" type we detected. The morphology on Bennu described as exfoliation corresponds to the "near-vertical spalling" type we observed. Another morphology of Bennu described as disaggregation of a clasts-bearing boulder, that leads to highly irregular, hummocky, profiles of some Bennu boulders, corresponds to the advection-like erosion we observed for the "breccia" type. These findings could imply that the morphologies on Bennu are formed by meteoroid bombardment and not thermal stresses or, perhaps less likely, that very different regimes of thermal stresses present on bodies of different dynamical and orbital properties (e.g., Molaro et al., 2017; Ravaji et al., 2019; Patzek and Rüsch, 2022) lead to the development of similar morphologies. It is also very likely that different processes lead to the development of similar morphologies thus limiting, in some cases, the inference based on geomorphology alone.

Another argument for the predominant role of thermal stresses in eroding boulders on Bennu is the preferential fracture orientation in planar view (Delbo et al., 2022). Since the preferentially oriented fractures formed by thermal stresses are more abundant than the randomly oriented fractures formed by impact, it is suggested that fragmentation driven by thermal stresses dominates (Delbo et al., 2022). Interestingly, a preferential orientation of fractures has been measured on the Moon as well (Ruesch et al., 2020) where the thermal stresses regime is very different (Molaro et al., 2017) and impact-fragmentation dominates (Hörz and Cintala, 1997; Hörz et al., 2020; Ruesch et al., 2020; Patzek and Rüsch, 2022). This type of fracturing with preferential orientation on the Moon is only one of several modes of failure (Fig. 8), and one of relatively little efficiency in terms of mass of the largest fragment to mass of the parent. Failure modes resulting in "core" or "cone" morphologies represent much more efficient modes. In summary, the observation made in Ruesch et al. (2020) and confirmed here, that most (>80%) of the boulders in a population of any age appear intact (in orbital images) and only few (<20%) are fractured, imply that the fragmentation process is characterized by a boulder weakening period when the boulder is internally damaged without disruption until sudden shattering occurs with a low $Alf/Ap$ ratio. During the weakening period the boulder accumulates internal damage due to multiple mild impacts (Rüsch et al., 2022) without displaying morphological signs of damage at the boulder scale. The length of this period is highly variable, increases with increasing boulder size, and can last for up to few hundreds million years for boulders >10 m in size. During this period the surficial properties of the boulder are modified by micrometeoroid bombardment (Rüsch and Wöhler, 2022) and, for petrologies susceptible to thermal stresses, by micro-cracking and -flacking (Patzek and Rüsch, 2022).

## Conclusion

The boulders located on the surface of the Moon are subject to high-velocity impact fragmentation and thus contribute to regolith formation, enabling the study of fragmentation and regolith development processes. Because of the heterogeneous distribution of boulders - and specifically fractured boulders - across the Moon's surface, we deployed a convolutional neural network to automatically identify and map them in LRO NAC images, effectively creating a global (60°N to S) catalog of fragmented boulders and other features with more than 180,000 entries. The dataset is available to the community for further studies. An analysis of the dataset allows us to draw the following, qualitative conclusions.

- Many high-velocity impact fragmentation morphologies can be described in terms of either concentric or radial flaws. In general, we find no obvious, qualitative correlation between the type of morphology and the selenographic position. However, regions of the same rock abundance in Diviner maps can be produced by rocks of different sizes and morphologies.
- New boulder morphologies are identified. Advection-like morphology is identified as a new type of rock degradation typical for breccia that differs from the more common abrasion-like morphology with debris apron of monolithic blocks. Impact morphologies for particular conditions (e.g., for a buried boulder, for fine-grained target containing boulders) studied in laboratory or numerical simulations for asteroids are identified for the first time on the Moon and can be studied with the dataset.
- Fragmented rock morphologies interpreted to be due to thermal stress on asteroid Bennu are identified, some for the first time, on the Moon, where the thermal stress regime is different. Therefore, some morphologies might be non diagnostic in terms of the fracture formation process. In addition, the presence alone of boulder scale morphological characteristics consistent with thermal stresses does not necessarily imply thermal stresses are the dominant fragmentation and erosion agent on the Moon and on asteroids.
- Our observations suggest that the fragmentation process on the Moon and possibly on asteroids is characterized by a weakening period with limited macroscopic morphological signs of damage at large scale until sudden catastrophic impact shattering occurs.

## Acknowledgments

O.R. is supported by a Sofja Kovelevskaja Award of the Alexander von Humboldt foundation. This work was supported by a Google Cloud Research grant.